\documentclass[aps,prl,epsfigure,subequations,showpacs,twocolumn]{revtex4}
\usepackage{dcolumn}
\usepackage{bm}
\usepackage{graphicx}
\usepackage{amsmath}
\usepackage{latexsym}
\usepackage{amsfonts}
\usepackage{amssymb}
\usepackage{array}
\usepackage{epsfig}
\usepackage{times}

\newcommand{\ket}[1]{\left\vert#1\right\rangle}
\newcommand{\miniket}[1]{\vert#1\rangle}

\newcommand{\minisprod}[2]{\langle#1\vert#2\rangle}

\newcommand{\minisand}[3]{\langle#1\vert#2\vert#3\rangle}

\begin{document}

\title{Nested entangled states for distributed quantum channels}

\author{C. Di Franco, M. Paternostro, and M. S. Kim}

\affiliation{School of Mathematics and Physics, Queen's University, Belfast BT7 1NN, United Kingdom}

\date{\today}

\begin{abstract}
We find a coupling-strength configuration for a linear chain of $N$ spins which gives rise to simultaneous multiple Bell states. We suggest a way such an interesting entanglement pattern can be used in order to distribute maximally entangled channels to remote locations and generate multipartite entanglement with a minimum-control approach. Our proposal thus provides a way to achieve the core resources in distributed information processing. The schemes we describe can be efficiently tested in chains of coupled cavities interacting with three-level atoms.
\end{abstract}


\maketitle

Two important topics can be placed at the forefront of current research on reliable devices for quantum information processing (QIP): the establishment of ``on demand" quantum channels among spatially distributed processors of any nature and the design of schemes that do not require the accurate addressing of the elements of a multipartite system. The first goal would allow the grounding of distributed QIP as an experimentally viable architecture for the manipulation of quantum information. Indeed, it is well-known that entangled channels between remote nodes of a device, together with the exchange of classical information, are pivotal resources in distributed QIP~\cite{distri}. Their realization has stimulated considerable experimental endeavors in many physical setups, from all-optical based to light-matter ones~\cite{kuzmich}. The second goal will constitute a major practical step forward in the process of narrowing the gap between experimental capabilities and theoretical requirements and is thus a crucial task to pursue. Very recently, it has been realized that effective control, communication and transfer of quantum information is achievable by means of engineered interactions that globally address a multi-spin system~\cite{spins,cambridge}. This has effectively paved the way to the use of multiple-spin couplings, requiring just a limited amount of external interventions, for the use of built-in collectively coupled systems. The factual combination of both these tasks would represent a key result not only for QIP but also, for instance, for the experimental simulation of quantum many-body dynamics: the design of less-demanding ways to create strong quantum correlations between spatially separated system will open up the field to the practical realization of quantum simulators.  

In this paper, we perform an important step in the design of resource-limited architectures of entangled channels and show that a quantum spin-chain can be exploited in order to  create and distribute multiple copies of maximally entangled states to remote locations of a delocalized system. This is possible by cleverly modifying the coupling configuration known to allow perfect state transfer in a linear spin-chain~\cite{cambridge}, a result achieved with the use of the recently introduced information-flux approach to multi-spin dynamics~\cite{informationflux}. Differently from the usual approach, this powerful tool lets us gather complete insight into the dynamics under such a modified coupling-pattern without relying on explicit analysis of the chain's spectrum. We reveal the generation, under minimal control conditions, of an unexpected nested structure of multiple Bell pairs and design simple schemes for their iterative extraction and the creation of multipartite entanglement of the Greenberger-Horne-Zeilinger (GHZ) form~\cite{ghz}. It is very important to stress that our scheme does not require any pre-built entangled resource. Entanglement (nested as well as multipartite) is generated in the chain solely as a result of spin-nonpreserving dynamics. The Hamiltonian model we need for our purposes can be simulated by the adiabatic interaction between coupled optical cavities, each embedding a  three-level atom in a $\Lambda$ configuration. This requires the introduction of spurious local magnetic fields whose effect is addressed for experimentally realistic parameters and shown to be only mildly affecting the effectiveness of our schemes.

We consider $N$ spins-$1/2$ particles in an open-chain configuration. Each spin is coupled to its nearest neighbors by the anisotropic and inhomogeneous $XY$ Hamiltonian $\hat{\cal H}=\sum_{i=1}^{N-1}(J_{X,i}\hat{X}_i\hat{X}_{i+1}+J_{Y,i}\hat{Y}_i\hat{Y}_{i+1})$. Here, $\hat{\sigma}_i$ (with $\sigma=X,Y,Z$) is the operator applying the $\sigma$-Pauli operator to spin $i$ and the identity to any other element of the chain. $J_{\sigma,i}$ is a pairwise coupling strength. For the sake of definiteness, we consider $N$ to be an odd number. It is well known that perfect quantum state transfer in a linear chain is possible for weighted coupling strengths following the pattern $J_{\sigma,i}=\lambda\sqrt{i(N-i)}$. After a time $t^*={\pi}/{\lambda}$, the state of the first qubit is perfectly transferred to the last one~\cite{cambridge}. Our protocol for the simultaneous generation of multiple pairs of Bell states without local control makes use of such a pattern in a novel way and exploits a useful technique for the analysis of spin-chain dynamics. In Ref.~\cite{informationflux} we introduce and analyze the concept of information flux in multiple-spin systems as the influences that the dynamics of a chosen spin receives from all the other elements of a register. Here, it is sufficient to state that the information flux between the $\hat{X}$ ($\hat{Y}$) operators of the first and last qubits in a chain engineered depends on an {\it alternate} set of coupling strengths. For instance, the information flux from $\hat{X}_1$ to $\hat{X}_N$ depends only on the set $\{J_{Y,1},J_{X,2},\!..,J_{Y,N-1}\}$ and is independent of any other coupling rate in the chain. This result is key to our investigation on how to generate a product of Bell states. Indeed, we observe that by suitably engineering the coupling pattern we can suppress the information flux between $\hat{Y}$ operators while keeping the flux between $\hat{X}$ operators unchanged (or vice versa). A simple way to obtain this is by setting $J_{\sigma,i}=0$ with $\sigma=X$ (Y) for odd (even) $i$. If the remaining couplings follow the given pattern with $\sigma=Y$ (X) for odd (even) $i$ it is straightforward to see that a unit information flux from $X_1$ to $X_N$ is achieved after an interaction time $t^*$.

\begin{figure}[t]
\psfig{figure=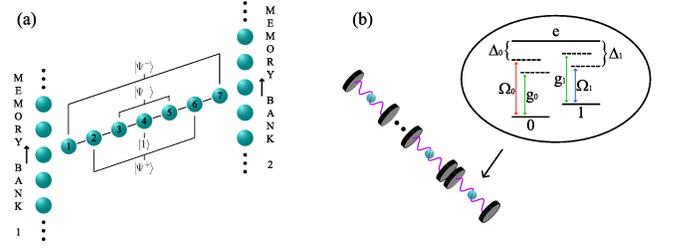,width=8.5cm}
\caption{{\bf (a)}: Pattern of multiple Bell states in a nested configuration (a {\it matryoshka state}) for a chain of $N=7$ spins. This structure is characteristic of the pattern of coupling rates that we propose. The scheme for the distribution of Bell states is also shown. The state of the boundary spins of the chain prepared in a matryoshka state is swapped with the state of a pair of systems belonging to two memory banks. {\bf (b)}: Scheme for the generation of matryoshka states: A chain of coupled cavities with embedded three-level $\Lambda$ atomic systems. The inset shows the coupling scheme for effective spin-spin Hamiltonian.}
\label{fig:matryoshka}
\end{figure}

In problems of multiple-spin dynamics, the analysis of the evolution of two-site correlation functions is often useful in order to understand the way (quantum as well as classical) correlations are arranged within a register. The information-flux approach that we explicitly use in this paper, being essentially performed in the Heisenberg picture, makes such an investigation particularly meaningful. We concentrate on the evolution of two-site operators which are symmetric with respect to the center of the chain, {\it i.e.} $\hat{X}_i\hat{X}_{N-i+1}$ and $\hat{Y}_i\hat{Y}_{N-i+1}$ (with $i=1,..,{(N-1)}/{2})$. In what follows, $\tilde{{O}}(t)$ stands for the time-evolved operator $\hat{O}$ in Heisenberg picture. After time $t^*$ for optimal state transfer, we have 
\begin{equation}
 \begin{split}
 &\tilde{X}_{i}(t^*)\tilde{X}_{N-i+1}(t^*)=(-1)^{(N-2i+1)/2}Z_{\bar{n}}\!\cdot\!\cdot Z_{N-\bar{m}+1},\\
 &\tilde{Y}_i(t^*)\tilde{Y}_{N-i+1}(t^*)=(-1)^{(N-2i+1)/2}Z_{\bar{m}}\!\cdot\!\cdot Z_{N-\bar{n}+1}
 \end{split}
\label{eq:evolution}
\end{equation}
with $2i-1=\bar{n}\!\pmod{2}$ and $2i=\bar{m}\!\pmod{2}$. We now reason along the lines of an information-flux approach: Eqs.~(\ref{eq:evolution}) show that, by preparing the initial state of the chain $\ket{\Psi(0)}$ in an eigenstate of the tensorial product of $\hat{Z}_{i}$ operators, there is a flux of information towards symmetric two-site spin operators. In particular, if $\ket{\Psi(0)}$ is a separable state, the form of Eqs.~(\ref{eq:evolution}) gives that $\minisand{\Psi(0)}{\tilde{\sigma}_i(t^*)\tilde{\sigma}_{N-i+1}(t^*)}{\Psi(0)}=\pm{1}$. By assuming $\ket{\Psi(0)}=\ket{00..0}_{12..N}$ with $\ket{0}_i$ ($\ket{1}_i$) the ground (excited) state of spin $i$ and going back to the Schr\"odinger picture, we realize that the only states satisfying the above conditions must have the structure
\begin{subequations}
\begin{equation}
\ket{\mu}=\ket{0}_c\otimes^{M}_{i=0}\ket{\Psi^+}_{2i+1,N-2i}\otimes^{M}_{i=1}\ket{\Psi^-}_{2i,N-2i+1},
\label{eq:statec0a}
\end{equation}
\vskip-0.75cm
\begin{equation}
\ket{\mu}=\ket{1}_c\otimes^{M}_{i=0}\ket{\Psi^-}_{2i+1,N-2i}\otimes^{M}_{i=1}\ket{\Psi^+}_{2i,N-2i+1}
\label{eq:statec0b}
\end{equation}
\label{eq:statec0}
\end{subequations}
with Eq.~(\ref{eq:statec0a}) [Eq.~(\ref{eq:statec0b})] valid for $c=1\!\pmod{2}$ [$c=0\!\pmod{2}$] and $c=(N+1)/{2}$ labelling the central site of the chain. We have called $M=(N-3)/4$ and $\ket{\Psi^\pm}_{pq}=({1}/{\sqrt{2}})(\ket{01}_{pq}\pm\ket{10}_{pq})$ are Bell states of spins $p$ and $q$. We thus end up with a {\it nested} product of alternate Bell pairs, the central spin being separable from the rest of the chain. We name such a peculiar configuration of entangled pairs at time $t^*$ a {\it matryoshka state} of the chain (matryoshka is Russian for "nesting doll"). An illustrative sketch of a matryoshka state of $N=7$ spins is presented in Fig.~\ref{fig:matryoshka} {\bf (a)}.
In general, matryoshka states (involving other Bell pairs) can be obtained every time the system is prepared in a separable state, with the spins in an eigenstate of $\hat{Z}_i$. This is a direct consequence of the particular form of the two-site operators. 

As we have already stated, the scenario where our investigation of matryoshka states is placed is distributed QIP. We have explicitly in mind architectures for processing devices which include remote nodes interconnected by channels of various nature. A mechanism able to distribute Bell pairs to two nodes of a processor is a necessary resource for the implementation of distributed QIP protocols~\cite{distri}. Here we show that the introduced matryoshka states play a useful role in such entanglement distribution process. Let us consider the chain studied so far as a bridging structure between two spatially separated processing nodes which we want to entangle. After having prepared an off-line matryoshka state of the bridging chain, we can connect the two boundary spins ({\it i.e.} spins $1$ and $N$) to two memory spins, labelled $m_1$ and $m_N$, each at one of the nodes. For definiteness, let us assume the memory spins to be prepared in  $\ket{00}_{m_1,m_N}$. The performance of swap gates within the $\{1,m_1\}$ and $\{N,m_N\}$ pairs of spins realizes the transformation [we concentrate on the case of $c=0\!\pmod{2}$, the other being essentially analogous] $\ket{\mu}\ket{00}_{m_1,m_N}\!\rightarrow\!\ket{010}_{1,c,N}\otimes^{M}_{i=1}\ket{\Psi^-}_{2i+1,N-2i}\otimes^{M}_{i=1}\ket{\Psi^+}_{2i,N-2i+1}$ while the memory qubits are now in $\ket{\Psi^-}_{m_1,m_N}$. However, it is often the case that a single entangled pair is not sufficient to the performance of a prefixed processing step~\cite{susana}. Alternatively, one can have situations where many entangled nodes have to be created, each node of a pair belonging to a different remote processor. We thus address the extraction of successive Bell states from the {\it reservoir} represented by a matryoshka state. A trivial solution would be to reset the system to its initial state $\ket{\Psi(0)}=\ket{00..0}_{12..N}$ (or, in general, to any separable state with the $i$-th spin in an eigenstate of $\hat{Z}_i,~\forall{i}$) and iterate the process. However, such a resetting step is unnecessary: After the first initialization of the bridging chain, all we need is only to operate on the boundary spins, following the minimal control conditions. In fact, if we do not reset the bridging chain after the swap gate and let it evolve again for a time $t^*$ under the same Hamiltonian $\hat{H}$ used for the generation of the matryoshka state, it is not difficult to see (either though an information-flux approach or by direct evolution of the state of the chain) that a Bell state of spins $1$ and $N$ is created, while the remaining part of the chain is in a separable state (in general different from $\ket{00..0}_{2,..,N-1}$ but having each spin in an eigenstate of $\hat{Z}$ anyway). By means of a second swap gate between $1$ and $N$ and two fresh memory spins (prepared in $\ket{00}$), we end up with the bridging chain in a fully separable state, eigenstate of the tensorial product of $\hat{Z}_{i}$'s. As we know, this state will evolve again in a matryoshka state. Clearly, by performing a sequence of swap gates with a time-interspacing equal to $t^*$, we can extract a Bell state at each round while the non-boundary ({\it i.e.} internal) spins of the chain will end up in a matryoshka or a separable state, in an alternate fashion. We have therefore realized a {\it conveyor belt} for Bell states. A sketch of the extraction scheme is provided in Fig.~\ref{fig:matryoshka} {\bf (a)}. Obviously, the ``storage'' of the extracted entanglement is an issue depending of the nature of the nodes constituting the distributed device linked by our chain. In fact, as an embodiment of the memory bank shown in Fig.~\ref{fig:matryoshka} {\bf (a)}, one can use long-lived qubits encoded, for instance, in the hyperfine spectrum of alkali-metal atoms or the nuclear spin of a nitrogen-vacancy defect in diamond. In this case, the storing qubit can be embedded in the end qubits of the chain (the electron spin being usable for the dynamics described in the paper). However, this is not the only possibility as the memory bank can also be embodied by a bosonic system (such as a cavity field, as in the setup that will be described later on). It is known that entanglement can be perfectly transferred from a two-qubit Bell state to two bosons prepared in coherent states and accumulation of more than a single ebit is possible (so that multiple extracted Bell pairs can be safely stored)~\cite{io}. Noticeably, the rate of entangled-pair extraction is bound by the assumptions of addressability at the end-spin of the chain. If we allow for local addressability of the qubits along the chain, it is clear that the generated entangled pairs are all available simultaneously, bringing the fan-out of our scheme to $(N-1)/2$ pairs.

The variety of possible applications offered by the pattern of coupling strengths addressed here is enriched by the possibility of generating a multipartite GHZ state. Indeed, let us slightly change the situation studied so far by considering the evolution of the bridging chain initially being prepared in $\ket{\Phi(0)}=\ket{11..1}_{12..N}$. By exploiting arguments entirely analogous to what has been done earlier in this work, we find the matryoshka state [here, we concentrate again on the case of $c=0\!\pmod{2}$] $\ket{\mu'}=\ket{0}_c\otimes^{M}_{i=0}\ket{\Psi^-}_{2i+1,N-2i}\otimes^{M}_{i=1}\ket{\Psi^+}_{2i,N-2i+1}$. The only difference between $\ket{\mu'}$ and $\ket{\mu}$ in Eq.~(\ref{eq:statec0b}) is the state of the central spin. The dynamics of the process guarantee that, if we let $\ket{\mu'}$ ($\ket{\mu}$) evolve again for a time interval $t^*$, we get $\ket{11..1}_{12..N}$ ($\ket{00..0}_{12..N}$). As a consequence, by preparing the (unnormalized) initial state $\ket{\mu}+\ket{\mu'}$, which accounts in preparing a matryoshka state and then {\it locally} acting on the central spin with a Hadamard gate (putting it into the superposition $\ket{0}_c+\ket{1}_c$), after a time $t^*$ the system evolves into the GHZ state $({1}/{\sqrt{2}})(\ket{00..0}+\ket{11..1})_{12..N}$. Such a resource is well known to be useful for multiagent protocols of distributed QIP such as quantum secret sharing, remote implementation of unknown operations and quantum average estimation~\cite{average}. A remark is due: our protocol lies entirely within the limited control framework where only a partial addressing of the chain is required. In this specific example, we only need to apply a single-qubit transformation at site $c$ before and after the {\it global} evolution ruled by $\hat{\cal H}$.

Very recently, arrays of coupled cavities interacting with atomic media  have been shown to hold great promises as flexible Hamiltonian simulators~\cite{simul,cavities}. Here we exploit such a potential in order to obtain our model $\hat{\cal H}$. We take advantage of the analysis conducted by Hartmann {\it et al.} in~\cite{cavities}, where an anisotropic $XY$ coupling has been obtained as an effective adiabatic Hamiltonian for a linear chain of optical cavities, mutually coupled via photon-hopping terms, each interacting with a three-level atom in a $\Lambda$ configuration.  The two ground states of each atom embody the computational space of each spin. The dipole-forbidden atomic transition between these states is realized as an adiabatic Raman transition through the excited state $\ket{e}_{i}$ ($i=1,..,N$). The cavity field drives off-resonantly the dipole-allowed channel $\ket{j}_i\leftrightarrow\ket{e}_i$ with Rabi frequency $g_j$ ($j=0,1$). A sketch of the physical setup and the coupling configuration within each atom-cavity system is given in Fig.~\ref{fig:matryoshka} {\bf (b)}. Two additional lasers are coupled to the atomic transitions with strength $\Omega_j$ and detuning $\Delta_j$. The coupling between two nearest-neighbor cavities is described by a tunnelinglike term with strength $J_C$~\cite{simul,cavities}. By assuming large detunings, both the excited level of each atom and the corresponding cavity field's degrees of freedom can be adiabatically eliminated so as to give an effective spin-spin Hamiltonian. The interaction between two effective spins is mediated via the excitation of virtual cavity photons which are coupled by the tunneling term.

The necessary anisotropy of the effective model is easily gathered by making the Rabi frequencies site dependent ({\it i.e.}, $\Omega_{0,i}$ and $\Omega_{1,i}$ are the Rabi frequency of the first and second lasers acting on the $i$th atom-cavity system). For the sake of simplicity, we consider lasers having the same frequencies $\omega_0$ and $\omega_1$. Moreover, any other physical parameter in the model is taken as homogenenous along the cavity-chain. By considering the possibility to differentiate all these parameters for each cavity, we have more degrees of freedom to fit the effective Hamiltonian to the required one. In what follows, in order to fix the ideas, we quantitatively address the case of $N=3$. In this specific case, the conditions to impose to the coupling strengths in order to create a matryoshka state can be reduced to $J_{Y,1}=J_{X,2}\ne0$ and $J_{X,1}=J_{Y,2}=0$. Each of these are expressed as adiabatic coupling strength, generalizing the expressions given in Ref.~\cite{cavities}. They are functions of the Rabi frequencies and relevant detunings appearing in the physical atom-cavity Hamiltonian model with photon hopping.

\begin{figure}[t]
\psfig{figure=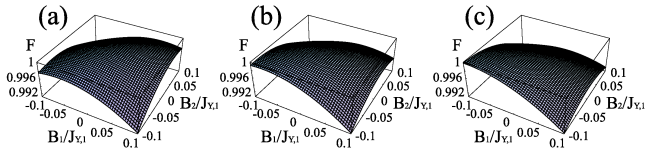,width=8.5cm}
\caption{Fidelity $F$ between the ideal three-spin matryoshka state and the state obtained with an Hamiltonian including additional local magnetic fields, against $B_1/J_{Y,1}$ and $B_2/J_{Y,1}$ for $B_3/J_{Y,1}=0,\,0.05$ and $0.1$ [{\bf (a)}, {\bf (b)} and {\bf (c)} respectively].}
\label{fig:fidelity}
\end{figure}

By means of a numerical analysis, we have obtained a suitable set of values~\cite{valori} giving spin-spin coupling strengths $J_{Y,1}=J_{X,2}=0.27$GHz and ${J}_{X,1}=J_{Y,2}={\cal O}(10^{-1})$MHz. This would allow us to write the Hamiltonian model between the effective spins along the chain $\hat{\cal H}=\sum_{i=1}^{2}(J_{X,i}\hat{X}_i\hat{X}_{i+1}+J_{Y,i}\hat{Y}_i\hat{Y}_{i+1})$. However, the a.c. Stark shifts experienced by the ground states of each atom as a result of the dispersive drivings give rise to additional magnetic-field-like terms $\sum_{i=1}^{3}B_i\hat{Z}_i$ with $B_1=7.8$MHz, $B_2=19.6$MHz, and $B_3=12.6$MHz. The nonzero (although small) value of these spurious local terms forces us to study the effects of additional magnetic fields in our model for matryoshka states generation. In Fig.~\ref{fig:fidelity} we have studied the fidelity $F=|\minisprod{\mu}{\mu^{(B)}}|$ against ${B_1}/{J_{Y,1}}$, ${B_2}/{J_{Y,1}}$ and for 3 different values of ${B_3}/{J_{Y,1}}$. Here, $\miniket{\mu^{(B)}}$ is the state that is obtained from the evolution of the initial state $\ket{000}_{123}$, for a time $t^*$, under the Hamiltonian $\hat{\cal H}+\sum_{i=1}^{3}B_i\hat{Z}_i$ while $\miniket{\mu}=\ket{0}_2\ket{\Psi^-}_{1,3}$ is the ideal three-site matryoshka state.
Evidently, for values of ${B_i}/{J_{Y,1}}$ ($i=1,2,3$) below $10\%$, which is the case in our numerical simulation, $F$ is always greater than 0.99. Therefore our model works properly also in the presence of small magnetic fields~\cite{commento}. We have evaluated the fidelity with the values obtained from our simulation, getting $F\simeq0.998$, which shows a striking resilience of our scheme to spurious local fields. On the other hand, it is worth stressing that the values we obtained for $J_{Y,1}$ and $J_{X,2}$ are well within the coherence time of the coupled cavity-atom system we consider. Indeed, the off-resonant nature of the couplings reduces the rate of both photon-loss and spontaneous emission from the excited atomic state $\ket{e}_i$. Indeed, photonic-crystal microcavities with embedded impurities are suitable canditates for implementation of our porposal as they are characterized by large cavity quality factors and small rates of spontaneous emission~\cite{yamamoto}.

We have introduced the matryoshka states, an entanglement structure in a linear chain of coupled spins resulting from a simple arrangement of the coupling pattern for perfect quantum state transfer. The multi-Bell nature of matryoshka states makes them suitable for the distribution of entangled channels in a delocalized quantum processor, which is one of the central requirements for reliable non-local QIP. Our protocol, lying entirely within a minimum control scenario, is so flexible that with just a single local operation, multipartite GHZ entanglement can be obtained. The interaction model we considered can be simulated in coupled-cavity chains with embedded three-level atoms, which can be used for testing at least the scheme for the generation of matryoshka states. However, the versatile nature of our proposal makes it appealing within the broad context of control-limited interacting many-body systems.

We thank G. M. Palma and F. Ciccarello for discussions. We acknowledge support from the U.K. EPSRC, The Leverhulme Trust (Grant No. ECF/40157) and QIPIRC.


\begin{thebibliography}{99}
\bibitem{commento} The study of the effects of additional local magnetic fields can be extended to a larger number of qubits (up to ten in our calculations), finding the same qualitative features we have described  and a good resilience of the protocol to values of $B_j/J_{\sigma,i}$ up to $10\%$. We notice that, for an alternate pattern of magnetic fields ($B_{2j}=-B_{2j+1}=B$), we have $F=1$ $\forall B$.
\end{thebibliography}
\end{document}